\documentclass{sig-alternate}

\usepackage{url}
\usepackage{paralist}
\usepackage{graphicx}
\usepackage[final]{pdfpages}
\usepackage{changebar}
\usepackage{graphicx}
\usepackage{balance}

\conferenceinfo{EASE}{'13, April 14-16 2013, Porto de Galinhas, Brazil}
\CopyrightYear{2013}
\crdata{978-1-4503-1848-8/13/04}

\begin{document}

\title{Naming the Pain in Requirements Engineering: \\ Design of a Global Family of Surveys and First Results from Germany}

\numberofauthors{2} 
\author{
\alignauthor
Daniel M\'{e}ndez Fern\'{a}ndez\\
       \affaddr{Technische Universit\"at M\"unchen}\\
%       \affaddr{Boltzmannstr. 3}\\
       \affaddr{Garching bei M{\"u}nchen, Germany}\\
       \email{mendezfe@in.tum.de}
% 2nd. author
\alignauthor
Stefan Wagner\\
       \affaddr{University of Stuttgart}\\
    %  \affaddr{Universit\"atsstrasse 38}\\
       \affaddr{Stuttgart, Germany}\\
       \email{stefan.wagner@informatik.uni-stuttgart.de}
}

\maketitle

\begin{abstract}
\textbf{Context:} For many years, we have observed industry struggling in defining a high quality requirements engineering (RE) and researchers trying to understand industrial expectations and problems. Although we are investigating the discipline with a plethora of empirical studies, those studies either concentrate on validating specific methods or on single companies or countries. Therefore, they allow only for limited empirical generalisations. \textbf{Objective:} To lay an empirical and generalisable foundation about the state of the practice in RE, we aim at a series of open and reproducible surveys that allow us to steer future research in a problem-driven manner. \textbf{Method:} We designed a globally distributed family of surveys in joint collaborations with different researchers from different countries. The instrument is based on an initial theory inferred from available studies. As a long-term goal, the survey will be regularly replicated to manifest a clear understanding on the status quo and practical needs in RE. In this paper, we present the design of the family of surveys and first results of its start in Germany. \textbf{Results:} Our first results contain responses from 30 German companies. The results are not yet generalisable, but already indicate several trends and problems. For instance, a commonly stated problem respondents see in their company standards are artefacts being underrepresented, and important problems they experience in their projects are incomplete and inconsistent requirements. \textbf{Conclusion:} The results suggest that the survey design and instrument are well-suited to be replicated and, thereby, to create a generalisable empirical basis of RE in practice. 
\end{abstract}


\section{Introduction}
Requirements engineering (RE) is a discipline that constitutes a holistic key to successful development projects as the elicitation, specification and validation of precise and stakeholder-appropriate requirements are critical determinants of quality~\cite{broy06_mbRE}. At the same time, RE is characterised by the involvement of interdisciplinary stakeholders and uncertainty as many things are not clear from the beginning of a project. Hence, RE is highly volatile and inherently complex by nature. 

Although the importance of a high quality RE and its continuos improvement has been recognised for many years, we can still observe industry struggling in defining and applying a high quality RE~\cite{MWLBC10}. The diversity of how RE is performed in various industrial environments, each having its particularities in the domains of application or the software process models used, dooms the discipline to be not only a process area difficult to improve, but also difficult to investigate for common practices and shortcomings. 

From a researcher's perspective, experimental research in RE thereby becomes a crucial and challenging task. It is crucial, as experimentation of any kind in RE, ranging from classical action research through observational studies to broad exploratory surveys, are fundamentally necessary to understand the practical needs and improvement goals in RE, to steer problem-driven research and to investigate the value of new RE methods via validation research~\cite{CDW12}. It is challenging, because we still need a solid empirical basis that allows for generalisations taking into account the human factors that influence the anyway hardly standardisable discipline like no other in software engineering. In consequence, qualitative research methods are gaining much attention~\cite{Seaman99}, and survey research has become an indispensable means to investigate RE.

\subsection{Problem Statement}
Although we are confident about the value of survey research to understand practical needs and to distill improvement goals in RE, we still lack a solid empirical survey basis. The reason seems to lie in an ironically paradoxical circumstance: The appropriate design of a survey in RE and the descriptive interpretation of the results going beyond purely observational, qualitative analyses and reasoning is very challenging, because we still lack empirically grounded theories in RE~\cite{CDW12}. In turn, we still lack such theories in RE as we still have no empirically sound survey basis. Available surveys in RE either investigate isolated techniques in application, or they focus on a small data population (single countries or companies) so that the findings of the surveys are hardly generalisable -- they cannot be viewed as representative. 

Yet missing is a series of empirical investigations of practical problems and needs in RE that allows for empirical generalisations to steer future research in a problem-driven manner.

\subsection{Research Objective}

As a long-term goal, we want to establish an open and generalisable set of empirical findings about practical problems and needs in RE that allows us to steer future research in a problem-driven manner. To this end, we aim at conducting a continuously and independently replicated, globally distributed survey on RE that investigates the state of the practice and trends including industrial expectations, status quo, experienced problems and how those problems manifest themselves in the process.

\subsection{Contribution}

We contribute the design of a globally distributed family of surveys on RE and first results from its start conducted in Germany. The survey will be replicated from 2013 on in a series of countries to manifest a clear understanding on the practical needs in RE and the inference of practically relevant improvement goals. 

\subsection{Outline}

The remainder of the paper is as follows. In Sect.~\ref{sec:RelatedWork}, we introduce available contributions in the context of our research, which gaps are left open, and how we intent to close those gaps. In Sect.~\ref{sec:Design}, we introduce the design of the family of surveys, including the research questions, the used methodology and instrumentation, an initial theory (to be extended during the replications), and the data analysis and validity procedures. The first results of the survey currently ongoing in Germany are presented in Sect.~\ref{sec:Results}, before giving a concluding discussion in Sect.~\ref{sec:Conclusion}.

\section{Related Work}
\label{sec:RelatedWork}

We directly focus on empirical investigations in our area and delimit from any philosophical discussions, opinion papers or solution proposals (see~\cite{WMMC05}) of which several valuable ones exist. We can classify related empirical investigations into two major areas: investigations of techniques and methods, and investigations of general practices and contemporary phenomena in industrial process environments. In both areas, we find survey research as well as technical action research among case and field studies. 

Contributions that investigate techniques and methods analyse, for example, selected requirements phases and which techniques are suitable to support typical tasks in those phases. Zowghi et al.~\cite{ZC05}, for example, conducted a survey about which techniques support the elicitation phase. A broader investigation of all phases is performed by Cox et al.~\cite{CNV09} who analysed the perceived value of the RE practices proposed by Sommerville and Sawyer. An exemplary survey on the choice of elicitation techniques is carried out by Carrizo et al.~\cite{CDJ08}. Studies like those reveal the effects of given techniques when applying them in practical contexts. 

Another type of studies on techniques and methods is often driven by the objective of investigating the improvement of specific variables when applying different techniques in same or similar contexts. For instance, we investigated the effects two different process models in RE had on the quality of the resulting artefacts (e.g.\ specification documents) by performing technical action research~\cite{MLPW11}. Abrah\~{a}o et al.~\cite{AICG11} raise the level of abstraction in this research area and set those kind of studies into the context of a framework that supports the validation of methods based on user perception while testing the framework with a family of experiments. In general, those studies give the opportunity to test the sensitivity of existing RE approaches in an industrial context, but they rely on a problem domain explored in advance and focus on pre-defined improvement goals. 

To reveal such industrial improvement goals and explore the problem domain to steer research activities, we mostly rely on field studies and surveys. One of the most commonly known survey is the Chaos Report of the Standish Group, examining, inter alia, root causes for project failures of which most ones are to be seen in RE, such as missing user involvement. Whereas the report is known to have serious flaws in its design negatively affecting the validity of the results~\cite{EV10}, other studies, such as the Success~study, conduct a similar investigation of German companies including a detailed and reproducible study design. Still, both surveys exclusively investigate failed projects and general causes at the level of overall processes. A similar focus, but exclusively set in the area of RE, had the study of Kamata et al.~\cite{IT07}. They analysed the criticality of the single parts of the IEEE software requirements specification Std.~830-1998 on project success. 

The focus of those studies, however, does not support investigation of contemporary phenomena and problems in industrial RE environments. Such investigations have, for example, been indirectly conducted by Damian et al.~\cite{DC06}. They analysed process improvements in RE and the relation to payoffs regarding, for example, productivity and the final product quality. Nikula et al.~\cite{nikula2000sps} present a survey on RE at organisational level of small and medium size companies in Finland. Based on their findings, they inferred improvement goals, e.g.\ on optimising knowledge transfer. 

A study investigating the industrial reluctance on software process improvement was performed by Staples et al.~\cite{SNJABR07}. They discovered different reasons why organisations do not adopt normative improvement solutions dictated by CMMI, such as no clear benefit and the relation to the company size. A field study with a broader data population of 60 cases in one company has been performed by Enam et al.~\cite{EM95}. They could infer recommendations to practitioners, such as the involvement of users in the elicitation process. A more curiosity-driven study to analyse typical project situations in companies was presented by us in~\cite{MWLBC10}. We could discover 31 project characteristics that directly influence RE. A survey that directly focused on discovering problems in practical settings was performed by Hall et al.~\cite{HBR02}. They empirically underpin the problems discussed by Hsia et al.~\cite{HDK93} and investigated a set of critical organisational and project-specific problems, such as communication problems, inappropriate skills or vague requirements, while those problems matched to a large extent with project characteristics we could discover.

\subsection*{Discussion} The previous non-exhaustive list of contributions reveals valuable observations when applying methods and techniques in sensitive, industrial contexts. Another introduced type of studies being directly related to our contribution comprehends surveys that focus on the industrial status quo and problems in RE. Although giving valuable insights into industrial environments, those studies do by now not allow for generalisation as they focus on single aspects in RE, such as problems in RE processes or RE improvements, or they focus on small subject populations (e.g.\ focus groups in single companies) and, thus, these studies by now remain not representative.

To close this gap in literature, we designed a family of surveys in joint collaboration with different researchers. The initial theory for the surveys relies on available study results as the ones introduced in the previous section and we expect it to change (along with the variables) over the years during replications due to an expected learning curve at us researchers. In addition, we present the first results from the survey conducted in Germany and illustrate a replication outline beginning from  2013 in different countries. By bringing together different interdisciplinary communities, the survey shall build an empirical basis for empirical generalisations and problem-driven research in RE.

\section{Design of a Family of Surveys}
\label{sec:Design}
In the following, we introduce the design of the family of surveys. Our overall long-term objective is to lay the empirical foundation to be continuously replicated in different countries. Those surveys aim at a generalisable investigation of the state of the practice and trends in RE including practitioners' expectations, the status quo in RE and its improvement, and contemporary problems experienced in companies. To support the dissemination of the results and the collaboration among the research communities, we provide a shared survey infrastructure relying on the same questionnaire, and we disclose the anonymised data to the PROMISE repository.

In the following, we formulate four research questions that build the frame for our surveys. Afterwards, we design the overall methodology and introduce the instrument, i.e.\ the (typed) questions and the categories, in Sect.~\ref{sec:Instrument}. In Sect.~\ref{sec:Theory}, we introduce an initial theory and our expectations we have gathered so far on basis of available literature, before concluding with the data analysis and validity procedures.

\subsection{Research Questions}
We formulate four research questions to steer the overall design of the surveys shown in Tab.~\ref{tab:rqs}.

\begin{table}[htb]
\caption{Research questions\label{tab:rqs}}
\begin{tabular}{p{0.1\linewidth}p{0.8\linewidth}}
\hline
\textbf{RQ~1} & What are the expectations on a good RE ? \\ 
\textbf{RQ~2} & How is RE defined, applied, and controlled? \\
\textbf{RQ~3} & How is RE continuously improved? \\
\textbf{RQ~4} & Which contemporary problems exist in RE and how do they manifest themselves in the process?\\
\hline
\end{tabular}
\end{table}

Not included in those research questions (but in the instrument) are questions to characterise the survey respondents and the industrial environment in which they are involved. The first research question aims at investigating the expectations and preferences the respondents have on a good RE. Research question~2 and~3 aim at investigating the status quo in the RE as it is established in their companies as well as industrial undertakings to continuously improve RE. Finally, the last research question aims at investigating which problems practitioners experience in their project environments, how these problems manifest themselves in the process, and to what extent those problems have already lead to failed projects. 

\subsection{Methodology}
\label{sec:Methodology}

\begin{figure}[hbt]
\centering
  \includegraphics[width=0.5\textwidth]{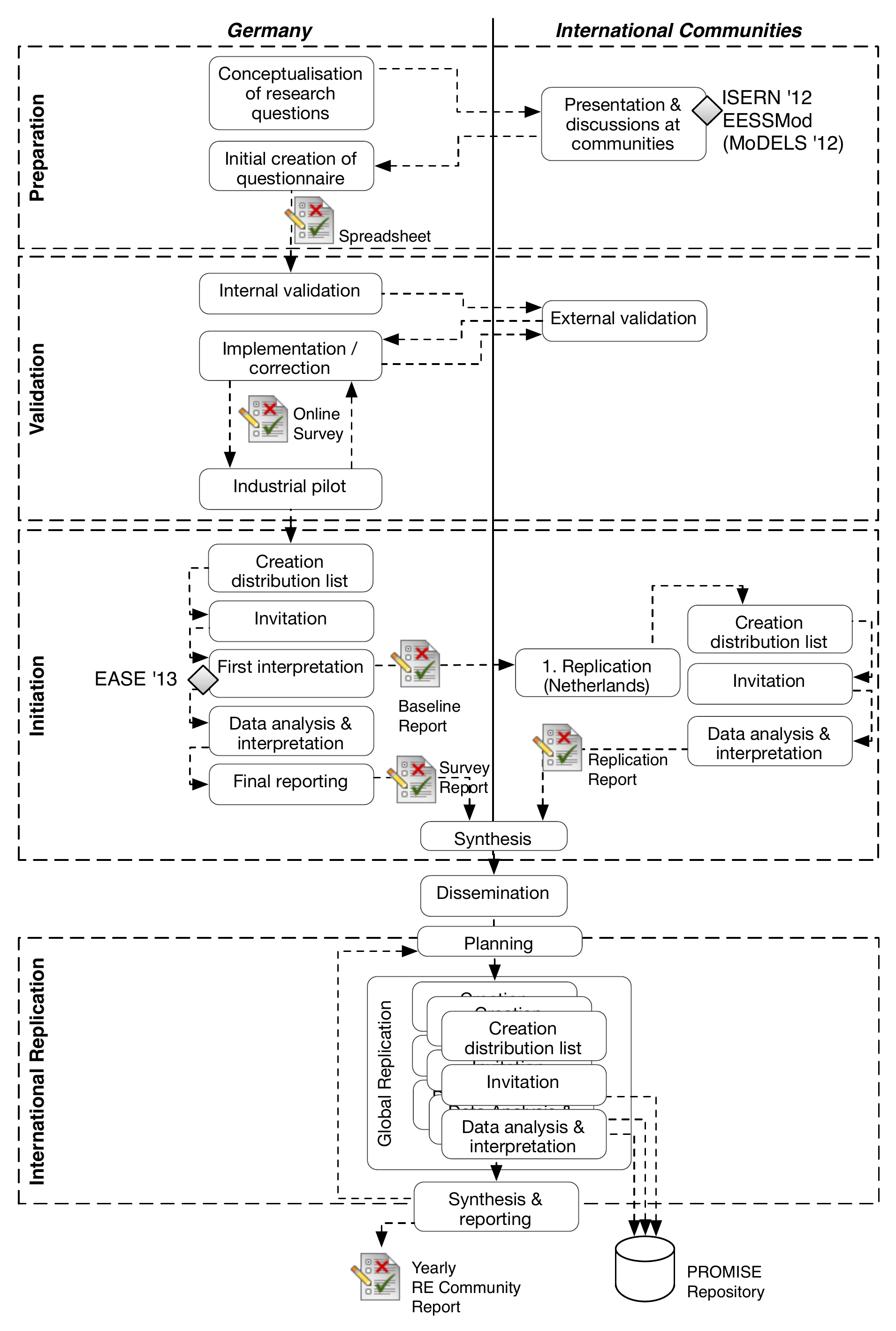}
  \caption{Overview of the methodology.}\label{fig.SurveyDesign}
\end{figure}

We designed our family of surveys based on experiences we made in academic research cooperations and discussions we follow in different international research and practitioners communities. This design contains four stages, which we illustrate in a simplified manner in Fig.~\ref{fig.SurveyDesign}. We distinguish between activities performed in isolation in Germany, and activities where we actively involved, or will involve again, international research communities. The first two stages comprehend the activities carried out to design and validate the survey structure (the research questions and the instrument). The third stage contains the survey implementation and the initial start in Germany from which we drew a baseline to report our findings in the paper at hand. The last stage comprehends the survey replications to be carried out from 2013 on. 

Considering the notion of ``replication'', we rely on the classification introduced by G\'{o}mez et al.~\cite{GJV10} and aim at empirical generalisations. Each replication of the survey is performed  independently by different researchers in different countries using the same infrastructure and instrument, before synthesising and reporting the overall results in joint collaboration. For the initial start of the survey, we rely on a coarse set of expectations for the definition of the research questions and of the instrument. Due to an expected learning curve on basis of the results, we  are aware that the theory will become more mature and change from year to year affecting the variables in the instrument.

In the following, we introduce the four stages of our methodology in more detail. The resulting instrument of the survey is introduced in the next Sect.~\ref{sec:Instrument}.

\subsubsection{Preparation}

Based on discussions at international events and the experiences we made during previous studies like the aforementioned ones (see Sect.~\ref{sec:RelatedWork}), we conceptualised an initial set of research questions and jointly discuss them at different community forums. The background and the thematic frame for the research questions was the topic of investigating the status quo in RE and its improvement in industry as well as contemporary problems practitioners have in their professional project setting to reason for improvement goals.  

We presented the idea of a joint survey at, for example, thematic workshops at the \emph{International Software Engineering Research Network} (ISERN) or at workshops like the \emph{International Workshop on Experiences and Empirical Studies in Software Modelling}, co-located with the \emph{International Conference on Model Driven Engineering Languages \& Systems}. For those preliminary discussions, we aimed at empirical research communities with a focus on RE rather than vice-versa. The reason is our own background and that we experienced discussions about more general principles of empirical designs of a family of surveys to be more effective in those communities taking into account the involvement of researchers and practitioners having a focus on RE. 

After checking for the resonance and initially agreeing informally with other researchers on a joint collaboration in the envisioned topic, we created an initial spreadsheet (the instrument) with a variety of questions and variables to answer our research questions. This questionnaire includes, where possible and reasonable, closed questions for a clear data analysis and to keep the effort low for practitioners when answering the questionnaire. To maximise the validity, we performed a series of validation tasks, which we introduce in the following.

\subsubsection{Validation}
\label{sec:Validation}
After creating an initial questionnaire, we performed a series of validation tasks, which took us in total three months. We first performed an internal validation of the questionnaire with a review by researchers not involved in the design of the questionnaire at the Technische Universit\"at M\"unchen and the University of Stuttgart. This internal validation should ensure, as a first step, that the closed questions are clearly interpretable and sufficiently complete w.r.t.\ the research questions, i.e.\ it should increase the internal and the construct validity. For the external review, we invited several researchers from different universities of which the ones listed in Tab.~\ref{tab:review} could do the review in the short time period (given the start of the winter terms). 

\begin{table}[htb]
\caption{Involved researchers
\label{tab:review}}
\begin{tabular}{p{0.27\linewidth}p{0.63\linewidth}}
\hline
\textbf{Review} & \textbf{Researcher}  \\
\hline 
Internal review & M. Broy, S. Eder, J. Eckhardt, K. Lochmann, J. Mund,  B. Penzenstadler \\ 
External review & M. Daneva, R. Wieringa (Twente)\\
& M. Genero (Castilla-La Mancha) \\
& J. M\"unch (Helsinki)\\
\hline
\end{tabular}
\end{table}

After increasing the construct validity with the external validation, we implemented the survey as a Web application using the \emph{Enterprise Feedback Suite}\footnote{The surveys, at the time of writing being password protected, can be reached at \url{RE-Survey.org}. This top level domain will also serve the hosting of the future replications.}. 

We conducted an industrial pilot phase with an industry participant. This participant has worked for five years as process consultant and has deep insights into the envisioned application domains, used RE standards, and he is familiar with the terminology used. His feedback and the analysis of the responses served to identify vague questions, incomplete answers in the closed questions, and how those answers apply to his context, thus, increasing the internal and the external validity. 

We complemented this pilot with two additional dry runs and external validations, before re-setting the data tables for the initiation of the survey.

\subsubsection{Initiation}

The initiation phase contains the survey conducted in Germany as well as its first replication conducted by Wieringa and Daneva in the Netherlands. Each of the surveys is closed and goes by invitation only to allow for a transparent and reproducible response rate and to ensure that the survey is answered by not more than one representative contact person per company (or business unit in case of large enterprises). In addition, the survey is anonymous due to the criticality of the questions (see also the next section). The replication in the Netherlands is triggered after drawing the first results we conclude from a baseline report -- presented in this paper. After the data analysis of the survey results (see Sect.~\ref{sec:DataAnalysis}), we will conduct a first synthesis and use the results to further disseminate the survey among the different research communities including the \emph{21st IEEE International Requirements Engineering Conference}, the \emph{19th Intl. Working Conference on Requirements Engineering: Foundations for Software Quality}, and the annual meeting of the \emph{International Software Engineering Research Network}. This dissemination also includes the detailed planning of the next iterations among different research sites of different countries, with which many we had already initial discussions during the preparation phase. 

In each survey round, we begin with the creation of a distribution list. This distribution list comprehends research partners from the universities while aiming at different roles from different companies of different sizes and application domains. Where possible, we inform the partners in advance and select, where reasonable, appropriate contact persons to support a high response rate. The official invitation to the survey contains
\begin{compactitem}
\item the basic information about the goals of the survey, the categories of questions and the context of the survey as part of a global family of surveys, and
\item the link to the survey and a password.
\end{compactitem}

We additionally ask the participants to forward the invitation, if necessary, and to inform us about the number of participants to support the inference of the response rate. We further re-ensure the participants about the anonymous nature of the survey and that they can add their e-mail address at the end of the questionnaire (not associated with the answers) so that we can inform them about the final results as an incentive. The survey in Germany was online from November 17th, 2012, until January, 31st, 2013. 

\subsubsection{International Replication}

We plan each subsequent replication to be performed in isolation by different researchers using the same (pre-agreed) questionnaire and survey infrastructure provided by us. Inherited from the nature of distributed survey replications, the replications will be performed, after a planning phase, independently and the survey design will change over the years due to a certain learning curve. To ensure a reproducible generalisation and the openness of the results to the communities, the anonymised results will be disclosed to the PROMISE repository\footnote{\url{ http://promisedata.googlecode.com}}. The overall aim is to establish a generalisable, open data basis to investigate industrial trends in RE.

\subsection{Survey Instrument}
\label{sec:Instrument}

Table~\ref{tab:instrument} summarises the questions of the survey in a simplified and condensed manner. We define in total 35 questions grouped according to the research questions and begin with a set of questions to characterise the respondents and the companies in which they work. At the end of the survey, the respondents can enter their e-mail address and freely add any other aspect that remained unaddressed in the survey.

\begin{table}[tb]
\tiny
\caption{Questions (simplified and condensed)}
\label{tab:instrument}
\begin{tabular}{p{0.03\linewidth}p{0.8\linewidth}p{0.1\linewidth}}
\hline
\textbf{RQ}& \textbf{Question} & \textbf{Type} \\ \hline 
-/-                   & What is the size of the enterprise? & Closed(SC)\\
                       & What is the main business area of your company? & Closed(MC)\\
                       & Does your company participate in globally distributed projects? & Closed(SC)\\
                       &  In which country are you personally located? & Open \\
                       &  In which application domain/branch are you most frequently involved in your projects? & Closed(MC)\\
                       & To which project role are you most frequently assigned to in those projects? & Closed(SC) \\
                       &  How would you classify your experience as part of this role? & Closed(SC)\\
                       &  Which organisational role takes your company usually in aforementioned projects? & Closed(SC)\\ \hline
RQ~1              &  How beneficial would you rate an improvement for following disciplines in your company?	 &  Likert scale \\
                       &  How challenging would you rate an improvement for following disciplines in your company?	 & Likert scale \\
                       &  Please rate the following statements on RE standardisation according to your expectations.			       & Likert scale \\
                       &  How important would you consider the following aspects when defining an RE standard?	& Likert scale \\
                       &  Which reasons do you agree with as a motivation to define an RE standard?	& Likert scale\\
                       &  Which reasons do you see as a barrier to define an RE standard? & Likert scale\\ \hline
RQ~2              &  Considering your regular projects, how would you classify you/your company to be involved in RE? & Closed(SC)\\
                       &  If you elicit requirements in your regular projects, how do you elicit them? & Closed(MC) \\
                       &  What RE standard have you established at your company? & Closed(Cond.) \\
                       &  Which of the following reasons apply to the definition of an RE standard in your company?	 & Closed(MC) \\
                       &  How would you rate the following statements to apply to your RE standard?				 & Likert scale \\
                       &  How is your change management defined regarding your RE? & Closed(MC)\\
                       &  Which of the following statements apply to the project-specific application of your RE standard? & Closed(MC)\\
                       &  How is your RE standard applied (tailored) in your regular projects?	 & Closed(MC) \\
                       &  How is the application of your RE standard controlled? & Closed(MC) \\ \hline
RQ~3              & Is your RE continuously improved? & Closed(Cond.) \\
                       &  What would you consider to be the motivation for a continuous improvement? & Closed(MC)\\
                       &  Which of the following statements applies regarding the continuous RE improvement?& Closed(MC)\\
                       &  Do you use a normative, external standard for your improvement? & Closed(SC) \\
                       & If you use an internal improvement standardad and not an external one, what where the reasons? & Open\\
                       & Which methods do you use for your RE improvement (regarding assessments/audits)? & Closed(MC)\\
                       & If you use metrics to assess your RE in the projects, which ones would you deem most important?& Open \\ \hline
RQ~4              &  Please rate the following statements for your RE standard according to your experiences. & Likert scale \\
                       &  How do the following (more general) problems in RE apply to your projects? & Likert scale\\
                       &  Considering your personally experienced problems (stated in the previous question), which ones would you classify as the five most critical ones (ordered by their relevance)? & Closed(Drop-down) \\
                       &  Considering your personally experienced most critical problems (selected in the previous question), how do these problems manifest themselves in the process, e.g.\ in requests for changes? & Open \\
                       &  Considering your personally experienced most critical problems (selected in the previous question), which would you classify as a major cause for project failures (if at all)? & Closed(MC) \\ \hline
\end{tabular}
\end{table}

For each question in the table, we denote whether it is an open question or a closed one and whether the answers are mutually exclusive single choice answers (SC) or multiple choice ones (MC). Most of the closed multiple choice questions include a free text option, e.g.\ ``other'' so that the respondents can express company-specific deviations from standards we ask for. We furthermore use Likert scales on an ordinal scale of 5 (e.g.\ ``strongly agree'' to ``strongly disagree'') to allow the respondents to explicitly select the middle when they have, for example, no opinion on the given answer options. Finally, we define conditional questions to guide through the survey by filtering subsequent question selection. For instance, if respondents state in RQ~2 that they have not defined any company-specific RE reference standard, the last questions of this section are omitted. For each of the questions (except the open ones), we define a series of answers, which we do not describe for reasons of space limitations in the table. Those answers can be taken from the results section~\ref{sec:Results}.

Finally, to define the answers in the closed questions, we establish a theory, which we introduce in the following. 

\subsection{Theory and Expectations}
\label{sec:Theory}

As stated in the previous sections, we define many questions on the basis of certain expectations we induce from literature and experiences. This applies to the definition of the questions and, in particular, of the answer possibilities in the closed questions. In the following, we introduce selected expectations we have. 

The first set of questions shown in Tab.~\ref{tab:instrument} serves to classify the study population, i.e.\ the participants involved and their experiences, as well as the company they represent. This allows us to analyse the relation of, for example, the company size, to the status quo in RE (see RQ~2 and~3). 

\textbf{Expectations on good RE (RQ~1).} The questions for RQ~1 shall initially characterise the expectations the respondents have on a good RE. In those questions, we directly ask for the expectations they have on the standardisation of RE as part of company-specific RE standards. We define different answer possibilities according to our experiences, e.g.\ concerning the expectations of the respondents on the standards; for instance, based on our investigation published in~\cite{MLPW11}, we expect respondents to demand for standards that focus on the RE artefacts with document templates rather than on strict processes and methods to allow for more flexibility and a better communication. We expect companies that are not aware of the RE artefacts to state in RQ~4 to have more problems with the completeness and consistency in the project-specific specification documents.

 We are also interested in the motivation and barriers the respondents expect when defining a company standard. Based on similar observations as in the previous questions, we expect respondents to see the improvement of the quality in the RE artefacts to be the main motivation for defining a company standard. Relying, for example, on the observations made by Nikula et al.~\cite{DC06}, we also suppose respondents to agree on the need of defining artefact models in the company standards to support knowledge transfer, because artefact models make implicit knowledge about the domain explicit (e.g.\ with templates or modelling guidelines). As a barrier to define a company standard, we rely on our experiences in research cooperations and expect respondents to agree on a higher process complexity and missing willingness for change, thus, we define the answer possibilities accordingly.

\textbf{Status Quo in RE (RQ~2).} Research question~2 serves to characterise the status quo in the RE of a company as well as the definition and application of their standard regarding tailoring. In general, we expect the standards to be rather immature compared to other disciplines due to the inherently complex nature of RE. We rely, for example, on the observations of Hall et al.~\cite{HBR02} and suppose the standards define coarse processes rather than well defined artefact models that support traceability. In consequence, we expect the application of that standard not to be mandatory while it is left to the expertise of project participants to tailor the standard at the beginning of a project. 

\textbf{Status Quo in RE improvement (RQ~3).} Regarding RQ~3, we rely on the observations made by Staples et al.\ in the area of software process improvement~\cite{SNJABR07}. We consequently expect especially small companies to not follow normative improvement approaches like CMMI. More general, we believe that normative improvement approaches are losing industrial attention as they are steered by goals and problems that do not necessarily match the ones of the companies. Therefore, we believe that qualitative, problem driven RE improvement approaches are gaining much attention and rely on the work of Petterson et al.\cite{PIGO08}.

\textbf{Contemporary problems in RE (RQ~4).} Finally, the last research questions aims at investigating contemporary problems the respondents see in their standard and the problems they experience in their projects. Regarding the problems in the standards, we rely again on the paradigms investigated in RQ~2, i.e.\ that the respondents see the problems in missing guidance to create syntactically complete and consistent artefacts. Regarding the investigation of which more general RE problems apply to their projects, how those problems manifest themselves in the process, and which of those problems were the major cause for project failures, we rely on a broad set of empirical investigations introduced in Sect.~\ref{sec:RelatedWork}. We define accordingly the list of problems and expect their selection as ordered in the following list:
\begin{compactenum}
\item Incomplete and/or hidden requirements
\item Inconsistent requirements
\item Terminological problems
\item Unclear responsibilities
\item Communication flaws within project teams and with customers
\item Moving targets (changing goals, business processes and/or requirements)
\item Technically unfeasible requirements
\item Stakeholders with difficulties in separating requirements from previously known solution designs
\item Underspecified requirements that are too abstract and allow for various interpretations
\item Unclear/unmeasurable non-functional requirements
\item Missing traceability
\item Weak access to customer needs and/or (internal) business information
\item Weak knowledge of customer's application domain
\item Weak relationship to customer
\item Time boxing/Not enough time in general
\item Discrepancy between high degree of innovation and need for formal acceptance of (potentially wrong/in\-complete/unknown) requirements
\item Volatile customer's business domain regarding, e.g.\ changing points of contact, business processes or requirements
\item ``Gold plating'' (implementation of features without corresponding requirements)
\item Insufficient support by project lead
\item Insufficient support by customer
\end{compactenum}

\subsection{Data Analysis}
\label{sec:DataAnalysis}

The data of the survey comprehends a mix of information about the companies and the RE standards used and expert opinions of the subjects involved in those companies. Moreover, the surveys do not rely on random samples as we opt for industry participants to whom we have contact, even if the participants distribute the invitation to further colleagues. Finally, regarding the expert opinions, we express the subjects' opinions with Likert scales, which are specified with ordinal scales with no interval data, i.e.\ the distances between the single values in the variables (e.g.\ ``strongly agree'', ``agree'', and ``disagree'') are not equally distributed. In other cases, we define the variables on purely nominal scales, e.g.\ the companies either apply certain methods for their RE improvement or they do not.

We apply descriptive statistics and use the mode and median for the central tendency of the ordinal data. To better understand the distribution of the data, we employ the median absolute deviations (MAD). For the nominal data, we calculate the share or respondents choosing the respective option. Although we do not expect significant results for current small sample size, we calculated the Kendall rank correlation for selected variables in the questionnaire. 

\subsection{Validity Procedures}
\label{sec:ValidityProcedures}

As a means to increase the validity of the family of surveys, we have built the instrument on basis of a theory induced from available studies (see Sect.~\ref{sec:Theory}). Furthermore, we conducted a self-contained, iterative validation phase, before initiating the first survey in Germany (see Sect.~\ref{sec:Validation}). In particular, we conducted internal reviews and external reviews to increase the internal and the construct validity via researcher triangulation. To support for the external validity in advance, we conducted a pilot phase in an industrial context and used the feedback in further external reviews and dry-runs of the surveys. The external validity, however, will eventually be supported during replications that finally support empirical generalisations.

\section{First Results from Germany}
\label{sec:Results}
In the following, we show the first results from the survey conducted in Germany. We invited in total 105 contacts to participate in the survey as representatives for their companies. In cases of large enterprises with different business units focusing each on different application domains, we invited for each business unit one representative (if known). The contacts arise from previous research cooperations or knowledge transfer workshops for practitioners hosted at the universities.

In the following, we first summarise the information about the study population, before describing the results for each of the research questions. Questions for which we have no sufficient data yet are omitted (mostly additional open answer possibilities in MC questions). We separate the descriptive statistics from initial interpretations. Where possible, we directly refer to the theory introduced in Sect.~\ref{sec:Theory}. 

\subsection{Study Population}

At the point of writing, we have 30 completed questionnaires. This gives us a current response rate of 29~\%, which we expect to increase as the questionnaire will be open until January 31, 2013. Most respondents (mode) work in an enterprise with more than 2,000 employees. The median are enterprises with
251--500 employees. Therefore, the respondents tend to work in larger companies, but we have representatives from companies of all sizes. The respondents represent a broad range of software 
domains (see Tab.~\ref{tab:study-population}).

\begin{table}[htp]
\caption{Study population's software domains}
\begin{tabular}{p{0.8\linewidth}p{0.1\linewidth}}%{lr}
\hline
\multicolumn{2}{l}{\textbf{Main business area}} \\
\hline
Custom software development & 23 \% \\
Standard software development & 37 \%\\
Project management consulting & 30 \% \\
Software process consulting & 27 \% \\
IT consulting & 33 \%\\
Embedded software development & 26 \%\\
\hline
\end{tabular}
\label{tab:study-population}
\end{table}%
Most of the respondents (67~\%) work in companies that participate in globally distributed projects. The large majority of respondents are located in Germany with a few exceptions located in Switzerland, Austria or France. 80~\% of the respondents are experts with more than three years of experience. The rest has 1--3 years of experience. The companies of the the respondents cover all the roles (customer, contractor, product development) in their projects. 23~\% state that they take the customer role, 43~\% take the role of a contractor and 37~\% refer to product development.

\subsection{Expectations on a Good RE (RQ~1)}
Regarding the practitioners' expectations on a good RE, we cover two topics: RE process improvement and expectations on (RE) company standards.

\textbf{Process Improvement.} We first looked at how improving the RE compares to (software) process improvements in other areas. Table~\ref{tab:beneficial-improvement} shows that the respondents considered process improvement in all offered areas as beneficial. 

\begin{table}[htp]
\caption{How beneficial would you personally rate an
improvement \ldots in your company? (Not beneficial
at all: 1 \ldots Very beneficial: 5)}
%\begin{center}
\begin{tabular}{p{0.55\linewidth}cp{0.08\linewidth}cp{0.03\linewidth}cp{0.03\linewidth}}
\hline
\textbf{Phase/discipline} & \textbf{Mode} & \textbf{Med.} & \textbf{MAD} \\
\hline
Requirements engineering & 5 & 5 & 1 \\
Project management & 4 & 4 & 1\\
Architecture and design & 4 & 4 & 1\\
Implementation & 4 & 4 & 1  \\
Quality assurance & 4 & 4 & 1\\
\hline
\end{tabular}
%\end{center}
\label{tab:beneficial-improvement}
\end{table}%

Only RE, however, was considered very beneficial. For all the results, the deviation was small. In addition, the respondents considered in high uniformity only RE improvements to be very challenging (Tab.~\ref{tab:challenge-improvement}).
Again, however, all disciplines were considered as more or less challenging. The deviation
here was also small.

\begin{table}[htp]
\caption{How challenging would you personally rate an
improvement \ldots in your company? (Not challenging
at all: 1 \ldots Very challenging: 5)}
%\begin{center}
\begin{tabular}{p{0.55\linewidth}cp{0.08\linewidth}cp{0.03\linewidth}cp{0.03\linewidth}}
\hline
\textbf{Phase/discipline} & \textbf{Mode} & \textbf{Med.} & \textbf{MAD} \\
\hline
Requirements engineering & 5 & 5 & 0 \\
Project management & 4 & 3 & 1\\
Architecture and design & 4 & 4&1 \\%3.5 & 0.5\\
Implementation & 4 & 3 & 1  \\
Quality assurance & 4 & 4 & 1\\
\hline
\end{tabular}
%\end{center}
\label{tab:challenge-improvement}
\end{table}%

\textbf{Requirements Engineering Standard.} We then asked about the opinion of the respondents on standards in RE. On average, we see an agreement or moderate agreement on most statements we offered: The standardisation of RE improves the overall process quality. Offering standardised
document templates and tool support benefits the communication and increases the
quality of the artefacts. The structure of documents should be standardised across
different project environments, but the process itself should be left open for project
participants. The only statement that received on average moderate disagreement was
that the standardisation of RE hampers the creativity. For all statements, the deviation
in the answers was low (MAD: 1).

Building on that, we asked about how important different aspects of a potential
company-specific standard reference model are. The results in Tab.~\ref{tab:important-aspects}
show that all the offered aspects seem to be rather important. 

\begin{table}[htp]
\caption{How important would you consider \ldots when defining a standard
RE model? (Not important at all: 1 \ldots Very important: 5)}
\begin{tabular}{p{0.55\linewidth}cp{0.08\linewidth}cp{0.03\linewidth}cp{0.03\linewidth}}
\hline
\textbf{Aspect} & \textbf{Mode} & \textbf{Med.} & \textbf{MAD} \\
\hline
Definition of artefacts & 5 & 4 & 1 \\
Definition of roles & 4 & 4 & 1\\
Definition of methods & 4 & 4 & 1\\
Tool support for V\&V & 3 & 3.5 & 0.5  \\
Support of impact analysis & 4 & 4 & 1\\
Process integration & 3 & 4 & 1\\
Support for agility & 5 & 4 & 1 \\
Support for prototyping & 4 & 4 & 1\\
\hline
\end{tabular}
\label{tab:important-aspects}
\end{table}%
The definition of artefacts and the support for agility received the highest number of \emph{Very important} answers. For tool support, for V\&V and for the deep integration with other phases, the most common answers were neutral. The deviation was again low with a MAD between 0.5 and 1.

When asked about the motivation for a company-wide reference model for
RE, the respondents agreed moderately with most of the given reasons
as shown in Tab.~\ref{tab:motivation-standard}. Exceptions are \emph{Better
quality assurance of artefacts} that received mostly agreements and
\emph{Formal prerequisite in my domain} that mostly received disagreement.
The deviation in all reasons was 1 or lower. 

\begin{table}[htp]
\caption{What do you agree with as a motivation for a reference model?
(I disagree: 1 \ldots I agree: 5)}
\begin{tabular}{p{0.55\linewidth}cp{0.08\linewidth}cp{0.03\linewidth}cp{0.03\linewidth}}
\hline
\textbf{Aspect} & \textbf{Mode} & \textbf{Med.} & \textbf{MAD} \\
\hline
Compliance to regulations & 4 & 3 & 1 \\
Seamless development & 4 & 4 & 1\\
Better tool support & 4 & 4 & 1\\
Prerequisite in domain & 1 & 2 & 1  \\
Support of distributed dev. & 4 & 3 & 1\\
Better progress control & 4 & 4 & 1\\
Better QA of artefacts & 5 & 4 & 1 \\
Support for benchmarks & 4 & 3 & 1\\
Support of project mgmt. & 4 & 4 & 0\\
Higher efficiency & 4 & 4 & 1 \\
Knowledge transfer & 4 & 4 & 1\\
\hline
\end{tabular}
\label{tab:motivation-standard}
\end{table}%

When asked about barriers to defining a company-wide reference
model for RE, the respondents mostly were on average neutral to
our proposed reasons (Tab.~\ref{tab:barrier-standard}). 
Only the missing willingness for change in the company was agreed by most
of the respondents. 

\begin{table}[htp]
\caption{What do you see as barrier for a reference model?
(I disagree: 1 \ldots I agree: 5)}
\begin{tabular}{p{0.55\linewidth}cp{0.08\linewidth}cp{0.03\linewidth}cp{0.03\linewidth}}
\hline
\textbf{Aspect} & \textbf{Mode} & \textbf{Med.} & \textbf{MAD} \\
\hline
Higher process complexity & 3 & 3 & 1 \\
Higher communication demand & 3 & 3 & 1\\
Lower efficiency & 3 & 3 & 1\\
Missing willingness for change & 5 & 4 & 1  \\
Missing poss. for standardisation & 3 & 3 & 1\\
\hline
\end{tabular}
\label{tab:barrier-standard}
\end{table}%

\subsubsection*{Interpretation} 

The respondents seem to see many potential
benefits in a RE reference model and RE improvement, but it is not a prerequisite in many
domains. The definition of artefacts and support for agility have a slightly higher importance in such a reference model showing the relevance of both topics. The main barrier against
such a model seems to be the general missing willingness to change. So far, those results seem to underpin our theory about the importance given to the artefacts and that the willingness to change barriers the establishment of an RE standard. We cannot, however, directly underpin the expected demand for knowledge transfer.

\subsection{Status Quo in RE (RQ~2)}

After the expectations, we asked the respondents how they are involved
in RE in regular projects. Most of the respondents (43~\%) elicit and specify the
requirements themselves. If they elicit requirements, we asked them about
how they elicit them. Of the respondents, 73~\% use workshops and discussions
with the stakeholders, 47~\% change requests, 43~\% prototyping, 40~\% agile
approaches at the customer's site and 13~\% other approaches.

Almost half of the respondents (43~\%) use an own RE reference model that 
defines the process including roles and responsibilities. A third of the respondents
have also an own RE reference model but one that defines the coarse process with
(coarse) artefacts, milestones and phases. A reference model that focuses on artefacts and 
offers document templates is in use at 30~\% of the respondents. A standard that is 
predefined by the development process (e.g.\ Rational Unified Process) employ 17~\%, 
and 13~\% use a standard that is predefined according to a regulation (e.g.\ ITIL). Only 13~\% 
use no RE reference model at all.

The main reason for the definition of an RE reference model were company-specific
demands (65~\%). Only 12~\% had an explicit demand from a customer and 8~\% because
of arguments from the sales department. Other reasons include \emph{to make requirements
more uniform} and \emph{quality and standardisation}.

Overall, the respondents rate their RE reference model well in terms of what it contains.
They mostly moderately agree with the statements about what their reference model
contains (Tab.~\ref{tab:statements-standard}). Only the weaker statement that the model
has a differentiated view on different classes of requirements but not their dependencies
is rated mostly as neutral. By looking at the MAD, however, we observe that the deviation
is, for all statements but the mentioned weaker one, high with 2.

\begin{table}[htp]
\caption{How would you rate \ldots to apply to your RE reference model?
(I disagree: 1 \ldots I agree: 5)}
\begin{tabular}{p{0.55\linewidth}cp{0.08\linewidth}cp{0.03\linewidth}cp{0.03\linewidth}}
\hline
\textbf{Aspect} & \textbf{Mode} & \textbf{Med.} & \textbf{MAD} \\
\hline
Relies on architect. model & 4 & 3,5 & 1,5 \\
Classes of reqs. \& dependencies & 4 & 4 & 2\\
Classes of reqs., no dependencies & 3 & 3 & 2\\
Tracing & 4 & 3 & 2  \\
Non-functional reqs. & 4 & 4 & 1 \\
\hline
\end{tabular}
\label{tab:statements-standard}
\end{table}%

The majority of the respondents (54~\%) agreed that in their company each
project can decide whether to use the RE reference model. That different business 
units have different standards as well as that all projects have to work according to the 
same standard each were agreed to by 27~\%.

The tailoring of the RE reference model is done with 57~\% of the respondents
at the beginning of the project by a project lead or a requirements engineer based
on experience. 31~\% have a tailoring approach that continuously guides the application
of the standard in their projects. 27~\% have tool support for tailoring their RE reference
model. Only 8~\% do not have a particular tailoring approach.

We found similar rates for how the application of the RE reference model is
controlled. 31~\% use project assessments, 39~\% analytical quality assurance, e.g.\ as
part of quality gates, and 39~\% constructive quality assurance, e.g.\ checklists or
templates. Almost a fifth of the respondents (19~\%) do not control the application
of their RE reference model. 

\subsubsection*{Interpretation} Almost half of the respondents use an own RE reference model with focus on coarse process descriptions with roles and responsibilities. 30~\% state to use a standard that defines typical RE artefacts. Regarding our theory, artefacts seem to remain underrepresented in the standards. The introduction of the reference models mostly came from inside the companies and were not forced on them by customers or standards. The respondents mostly see many features in their reference models but the deviation is high and, hence, the picture is more differentiated.
We also have many disagreements, which suggests that the used references models
are also highly different. In the vast majority of companies, there seems to be the opinion
that RE reference models need to be tailored while this tailoring is done at the beginning based on experiences, thus, confirming our theory. The concrete application is often not
controlled or controlled with very different means.

\subsection{Status Quo in RE Improvement (RQ~3)}

Most of the respondents (70~\%) employ continuous improvement to RE.  When asked about
the motivation about this continuous improvement, of those, 77~\% think
that this continuous improvement helps them to determine their strengths and weaknesses and
to act accordingly. 41~\% agree that an improvement is expected by their customers. For only
5~\% of those with continuous improvement, it is demanded by a regulation (e.g.\ CMMI, Cobit
or ITIL). 

We then asked about how they conduct their RE improvements. 64~\% systematically improve RE
via an own business unit or role. 9~\% improve RE via an external consultant. 22~\% do not
systematically improve RE, but it remains the responsibility of the project participants.  Other
mentioned means to systematical improvements are an internal task force, retrospectives and
company-wide open space events.

Most of the respondents with a continuous RE improvement (77~\%) do not use a normative,
external standard for their improvement. Several respondents use internal standards like an
internal process description system or best practices from literature.

Regarding the methods, 77~\% of the respondents that employ improvement qualitatively analyse their projects, e.g.\ with interviews to gather lessons learnt.  27~\% refer to particular metrics and measurements to automatically assess their projects. 

\subsubsection*{Interpretation}
Continuous improvement in RE seems to be performed in the majority of the companies. The
improvement is mostly driven from inside the companies and not from external standards or imposed
by customers or regulations. The improvement is achieved mostly by meetings, discussions
and interviews to understand the lessons learnt. The results seem to confirm our theory that normative improvement approaches like CMMI are losing attention as the respondents rely on qualitative, problem-driven improvement methods. We could not find, however, a strong or significant correlation between the company size and the choice for qualitative improvement methods.

\subsection{Contemporary Problems in RE (RQ~4)}

Finally, after laying the groundwork about how RE is defined, lived and improved, we wanted to understand current problems in RE in practice. First, we asked about problems with RE standards.
Of our offered problems, most of them had a strong agreement among the respondents that they are not a problem (Tab.~\ref{tab:problems-standard}). Only two problems have as most frequent answer
\emph{I agree}: \emph{\ldots gives no guidance on how to create the specifications documents} and \emph{\ldots
is not sufficiently integrated into risk management}. Both have lower medians, however, and the latter
problem also has a high deviation. We also observed that for \emph{\ldots does not sufficiently define a clear
terminology}, there is a higher median and also a bit increased deviation. Apart from these, the deviations
are low for all problems.

\begin{table}[htp]
\caption{Please rate \ldots for your RE reference model?
(I disagree: 1 \ldots I agree: 5)}
\begin{tabular}{p{0.55\linewidth}cp{0.08\linewidth}cp{0.03\linewidth}cp{0.03\linewidth}}
\hline
\textbf{Statement} & \textbf{Mode} & \textbf{Med.} & \textbf{MAD} \\
\hline
Is too hard to understand & 1 & 2 & 1 \\
Is too complex & 1 & 2 & 1 \\
Is too abstract & 1 & 2 & 1 \\
Doesn't support precise spec. & 1 & 2 & 1 \\
Doesn't scale  & 1 & 1.5 & 0.5 \\
Is too heavy weight & 1 & 2 & 1 \\
Is not flexible enough & 2 & 2 & 1 \\
Has no clear terminology & 1 & 3 & 2 \\
Doesn't guide to create artefacts & 4 & 2 & 1 \\
Doesn't allow for deviations & 1 & 2 & 1 \\
Doesn't define roles & 1 & 2 & 1 \\
Not integrated into proj.~mgmt. & 1 & 2 & 1 \\
Not integrated into design & 2 & 2  & 1 \\
Not integrated into risk mgmt. & 4 & 3 & 2 \\
Not integrated into test mgmt. & 1 & 2 & 1 \\
\hline
\end{tabular}
\label{tab:problems-standard}
\end{table}%

Second, we asked about more general problems in RE in the respondents' projects (Tab.~\ref{tab:problems-general}).
\begin{table}[htp]
\caption{How does \ldots apply to your projects?
(I disagree: 1 \ldots I agree: 5)}
\begin{tabular}{p{0.55\linewidth}cp{0.08\linewidth}cp{0.03\linewidth}cp{0.03\linewidth}}
\hline
\textbf{Problem} & \textbf{Mode} & \textbf{Med.} & \textbf{MAD} \\
\hline
Com.~flaws within the team & 4 & 3 & 1 \\
Com.~flaws w/ the customer & 4 & 4 & 1\\
Terminological problem & 3 & 3 & 1\\
Unclear responsibilities & 3 & 3 & 1  \\
Incomplete and/or hidden reqs. & 4 & 4 & 1\\
Insufficient support by proj. lead & 3 & 3 & 1 \\
Insufficient support by customer & 3 & 3 & 1 \\
Separating reqs. from solution & 3 & 3 & 1 \\
Inconsistent reqs. & 4 & 4 & 1 \\
Missing traceability & 3 & 3 & 1 \\
Moving targets & 2 & 3.5 & 1.5 \\
Gold plating & 2 & 3 & 1 \\
Weak access to customer needs & 3 & 3  & 1 \\
Weak knowl. of app. domain & 1 & 2 & 1 \\
Weak relationship to customer & 1 & 2.5 & 1.5 \\
Time boxing/not enough time & 4 & 4 & 1 \\
High degree of innovation & 3 & 3 & 1 \\
Technically unfeasible reqs. & 2 & 2 & 1 \\
Underspecified reqs. & 4 & 3.5 & 1.5 \\
Unclear non-functional reqs. & 3 & 3 & 1 \\
Volatile customer business & 3 & 3 & 1 \\
\hline
\end{tabular}
\label{tab:problems-general}
\end{table}%

There, we received a more mixed picture of RE in practice. On the one hand, six of the 
problems we offered were moderately agreed with by most of the respondents: communication flaws within the team and with the customer, incomplete and/or hidden requirements, inconsistent
requirements, time boxing/not enough time and underspecified requirements that are too abstract and allow for various interpretations. On the other hand, two problems were disagreed with: weak knowledge of the customer's application domain and weak relationship to customer. The rest was considered mostly neutral or was moderately disagreed with. The deviations were mostly small (1). In four problems, we
have a deviation of 1.5, which suggests a slightly higher diversity in the answers. Accordingly, we could not find any large or significant correlation between the problems stated by the respondents and the answers selected in RQ~1 and 2 (e.g.\ the relevance given to the artefacts and problems experienced w.r.t., inconsistency).

Finally, we asked the respondents to rank the problems they have experienced
according to their criticality. The most often mentioned problem in this ranking are
incomplete and/or hidden requirements. Also mentioned often are time boxing/not
enough time and inconsistent requirements. The most frequent answer for how those selected problems manifest themselves in the process was change requests and additional effort (e.g.\ for meetings). When asked which of the selected problems they saw as a major reason for experienced project failure, the highest answer was given to incomplete requirements (8 times), followed by communication flaws with the customer (6) and and moving targets as well as time boxing each being selected 5 times. The results on project failures, however, do not yet allow for a clear interpretation as not all respondents selected problems.

\subsubsection*{Interpretation} The major problem with standards seems to be that they do not guide the requirements engineers enough how to create specification documents. Also the integration with risk management is sometimes a problem but the deviations are higher and, hence, it seems to be different in different companies. The unclear terminology also receives some minor complaints. The picture is more mixed for general RE problems, but there are also no overwhelmingly large problems. The classic RE problems, communication, incomplete requirements, inconsistent requirements and not enough time seem to dominate. The application domain and the relationship to customers are in most companies not problematic.	

\section{Conclusion}
\label{sec:Conclusion}

In this paper, we contributed the design of a global family of surveys to overcome the problem of by now isolated investigations in RE that are not yet representative. With the family of surveys, we aim at establishing an empirically sound basis for understanding practical trends and problems in RE and for inferring representative improvement goals. Hence, the family of RE surveys will build a continuous and generalisable empirical basis for problem-driven research. 

The family of surveys relies on an initial theory induced from literature, and it is designed in joint collaboration with different researchers from different countries. An additional pilot phase rounded out the validity procedures we could perform in advance. We presented the first results from the survey conducted in Germany where 30 respondents of different companies participated (with an answer rate of 29\%) and showed initial trends in RE as well as problems the respondents experience in their practical settings. The first replication of the survey is currently performed in the Netherlands and the synthesis of both surveys will be further disseminated to bring together the various empirical and RE-specific research communities. Further replications are plan\-ned from 2013 on by the researchers that already committed themselves to our undertaking. Each replication uses the same infrastructure and is based on the same questionnaire while the interpretation of the results will be performed independently. To guarantee a reproducible design of the survey and of the results, we commit ourselves to disclose the anonymised data of each replication to the PROMISE repository. The overall objective is to establish a regularly performed survey replication to continuously adapt and, finally, manifest a theory on the practical status quo and problems in RE. 

\subsection{Relation to existing Evidence}
Based on our interpretations of the results w.r.t.\ the theory presented in Sect.~\ref{sec:Theory}, we could confirm selected empirical findings available in literature. For example, our results indicate that artefacts still remain underrepresented in available company standards for RE while the practitioners rate the definition of artefacts structures and contents in those standards to be important. Furthermore, we could already confirm the general reluctance of practitioners against available normative RE improvement standards, such as ones based on CMMI, which put the focus on assessing RE standards against pre-defined processes and methods. 

Regarding the problems observed in the respondents' pro\-ject environments, we could rank the following as the most problematic ones: incomplete and inconsistent requirements, communication flaws within teams and with customers, and time boxing. While the first four ones relate to existing evidence, the last one is, to us, remarkable.

\subsection{Impact/Implications}
We can directly derive two implications from our contributions. First, the results confirm a first coarse theory we draw on the basis of different isolated studies. This already allows researchers to steer their problem-driven research, i.e.\ they can define improvement goals on basis of a survey that already goes beyond isolated investigations and validation research; for instance, in the area of qualitative RE improvement methods and/or in the area of artefact orientation. Second, the family of surveys is and will remain open. This allows not only researchers to reproduce the results and their interpretation, but also practitioners to evaluate their own RE situation against overall industrial trends.

\subsection{Limitations}
Although we have first results from the survey conducted in Germany and can be confident about the next replications, we are aware that the design has still limitations. Most importantly and as already discussed, the initial theory is based on available contributions that investigate different aspects in RE in an isolated manner. Hence, the theory still needs to evolve and mature along with the variables in the instrument over the years during replications due to expected learning curves at us researchers, before we can finally establish a reliable and empirically grounded theory.

\vspace{3em}
\subsection{Future Work}

We plan the further coordination of the replications, their synthesis and dissemination as future work. The dissemination comprehends both empirical and RE research communities to support a variety of conceptual work in RE on basis of empirical sound findings. To this end, we cordially invite further researchers to join in for further replications over the next years to establish a generalisable empirical  basis on the state of the practice in requirements engineering.

\subsection{Acknowledgments}
We want to thank all participants of ISERN and EESSMod for fruitful discussions. In particular, we want to thank Silvia Abrah\~{a}o, Barry Boehm, Giovanni Cantone, Michael Chaudron, Maya Daneva, Marcela Genero, Lars Pareto, and Roel Wieringa for participating on the thematic workshops and giving us valuable feedback on the basic design of the family of surveys. Furthermore, we want to thank the internal reviewers listed in Tab.~\ref{tab:review} for their effort spent, and the external reviewers for their commitment to the whole undertaking. Finally, we are grateful to Michael Mlynarski for supporting us in the pilot phase and to our project partners for their participation.

\balance

\end{document}